\documentclass{cup-hpl}
\usepackage{graphicx}
\usepackage{epstopdf}
\usepackage{svg}
\usepackage{pgf}
\usepackage{siunitx}
\usepackage{mathtools}
\usepackage{breqn}
\usepackage{hyperref}
\usepackage{blindtext}
\usepackage{mathtools}
\usepackage{siunitx}
\usepackage{rotating}
\usepackage{array,booktabs}
\usepackage{ulem}
\usepackage{color}

\renewcommand{\sout}[1]{}
\usepackage{caption}

\begin{document}

\newtheorem{theorem}{Theorem}

\shorttitle{Enhanced dynamic range spatio-spectral metrology of few-cycle laser pulses}
\shortauthor{C. Alexe, A. Liberman, et al.}

\title{Enhanced dynamic range spatio-spectral metrology of few-cycle laser pulses}

\author[1,2,3,*]{Cristian Alexe}
\author[4,*]{Aaron Liberman}\corresp{ aaronrafael.liberman@weizmann.ac.il\\}
\author[5]{Saga Westerberg}
\author[5]{Andrea Angella}
\author[4]{Anda-Maria Talposi}
\author[5]{Erik Löfquist}
\author[1,2,3]{Alice Dumitru} 
\author[1,2,3]{Andrew H. Okukura} 
\author[5]{Flanish D’Souza}
\author[5]{Cornelia Gustafsson}
%\author[1]{Dan Gh. Matei}
\author[5]{Anders Persson}
\author[5]{Chen Guo}
\author[5]{Cord Arnold}
\author[5]{Olle Lundh}
\author[1,4]{Victor Malka}
\author[1,2,3]{Daniel Ursescu} \corresp{daniel.ursescu@eli-np.ro}

\address[1]{Extreme Light Infrastructure - Nuclear Physics, National Institute for Physics and Nuclear Engineering Horia Hulubei, 30 Reactorului str., M\u{a}gurele, 077125, Ilfov, Romania}
\address[2]{Doctoral School of Physics, University of Bucharest, Atomistilor 405, M\u{a}gurele, 077125, Ilfov, Romania}
\address[3]{Engineering and Applications of Lasers and Accelerators Doctoral School (SDIALA), National University of Science and Technology Politehnica of Bucharest, 313 Splaiul Independentei St., Bucharest RO-060042, Romania}
\address[4]{Department of Physics of Complex Systems, Weizmann Institute of Science, Rehovot 7610001, Israel}
\address[5]{Department of Physics, Lund University, P.O. Box 118, SE-22100 Lund, Sweden}
\address[*]{These authors contributed equally to this work}

\begin{abstract}
 Accurate spatio-temporal and spatio-spectral metrology is critical to the characterization and use of ultra-short, high-power lasers. The emergence of few cycle pulses, with bandwidths of tens or hundreds of nanometers, poses a significant challenge to existing metrology techniques. This is due both to large discrepancies in the sensitivities of the measurements at different wavelengths and to variation in the spectral intensity at those wavelengths. In this paper, the authors propose spectral filtering and stitching of the measurements as a robust, simple solution that enhances the dynamic range of the measurements, allowing accurate few-cycle pulse reconstruction. This enhancement is demonstrated using INSIGHT – the most commonly used spatio-spectral measurement device – as well as using IMPALA and spatially resolved Fourier transform spectrometry.

\end{abstract}

\keywords{ spatio-temporal characterization;  ultrashort laser pulses; few-cycle laser pulses}

\maketitle

\section{Introduction}
Ultrashort laser pulses are critical to many applications ranging from next generation medical technologies \cite{Glinec_Med_Phys_2006}, to non-destructive material testing \cite{Glinec_PRL_2005}, and to next-generation particle acceleration \cite{Tajima_PRL_1979}, to list a few. Since the invention of chirped-pulse amplification (CPA) \cite{Strickland_OpticsCom_1985}, femtosecond, multi-terrawatt systems have proliferated, enabling the exploration of extreme light-matter interactions \cite{mourou-rmp06}. These laser pulses have significant spectral bandwidths and have large apertures in the near-field. These characteristics together enable the emergence of spatially non-uniform pulse-front delays and pulse-durations from dispersion effects accumulated throughout the beam path \cite{Akturk_2010}. These spatially-dependent chromatic distortions, known as spatio-temporal couplings (STCs) have been found to be caused by numerous components throughout the laser generation, amplification, and beam transport, and by imperfect compensation of the STCs introduced in the stretcher-compressor CPA setups, and have been detected in a wide array of systems \cite{Jeandet:22}. 

Accurate spatio-temporal or spatio-spectral metrology and control is of great importance to the effective use of such laser-systems \cite{Walmsley_09}. In typical experimental setups, the goal is to avoid undesirable STCs, which can decrease intensity at focus and distort the quality of the beam \cite{Akturk_2010}. Recently, advanced schemes have been suggested for utilizing spatio-temporal manipulation of the beam \cite{Piccardo:25}. These include combining STCs with specialized optics that allow for an extended focus that propagates at a tunable velocity. The tunability of such flying-focus pulses has been experimentally demonstrated using several configurations \cite{Sainte-Marie_Optica_2017,Liberman_OL_2024,Pigeon_OE_2024} and shows promise in achieving higher energy particle acceleration \cite{Debus_2019_PRX,liberman2025electronaccelerationtunablevelocitylaser,Caizergues_NP_2020,Liberman_NatureCommunications_2025,Palastro_PRL_2020,liberman2025probingflyingfocuswakefields}, more efficient terahertz generation \cite{Simpson_SciRep_2024}, laser-wakefield acceleration of ions \cite{Zheng_PRL_2024}, and enabling laser-based undulators for X-ray free electron lasers \cite{Ramsey_CommPhys_2025}. Another use which has generated significant interest is the generation of vortex beams \cite{Sueda:04,Shi_China_2024}, which have shown promise in both electron acceleration \cite{Blackman_CommPhys_2022} and proton acceleration \cite{Willim_PRR_2023} and have been adapted to large-aperture, high-intensity beams \cite{Iancu_HPLSE_2024}. 

A number of spatio-spectral and spatio-temporal metrology methods have been developed \cite{Akturk_2010,Dorrer_2019,Jolly_IOP_2020}. Several are based on the use of spatially resolved Fourier transform spectrometry (SRFTS) \cite{Miranda_OL_2014,Pariente_NatPhot_2016}, using a spatially-filtered self-reference pulse. Perhaps the most widely used measurement method, INSIGHT \cite{Borot_OE_2018}, is based on a similar scanning interferometric setup, but the measurement is conducted at focus rather than in the near-field, and the Gerchberg–Saxton (GS) iterative algorithm \cite{GS_1972} is used to reconstruct the spectrally-resolved spatial profile of the beam. Other methods rely on spatially-resolved spectral interferometry \cite{STARFISH}, wavelength-resolved wavefront measurement techniques \cite{HAMSTER}, digital holography \cite{STRIPEDFISH}, optical filtering of the measured pulse \cite{Weise_2023}, broadband Young's Double Slit Interferometry \cite{Smartsev_JOPT_2022}, and hyperspectral imaging \cite{BATEY_UM_2014}. Some techniques use imaging spectrometers to measure the pulse, but this is limited to one dimension \cite{Witting:11} rather than fully measuring the pulse. Lately, a number of one-shot techniques have emerged, promising shot-to-shot measurement of the pulse \cite{Tang_LSA_2022,Smartsev_OL_2024,Howard_NaturePhotonics_2025,Kim_OE_2021}. These include IMPALA \cite{Smartsev_OL_2024}, which algorithmically separates out spectral information in the Fourier plane, allowing for multiple GS algorithms to be run for different colors.

In recent years, a number of systems have emerged that provide few cycle pulses with durations at or below 10 fs and with peak powers in the terawatt scale such as the Salle Noire laser at Laboratoire d’Optique Appliqué \cite{Guenot_NaturePhotonics_2017}, the PFS laser system at Ludwig-Maximilians-Universität München \cite{Nubbemeyer:17}, the LUCID laser system at the Lund Laser Centre (LLC) \cite{Balciunas:24}, and a number of facilities inside of ELI-ALPS \cite{Budriunas:17}. Some of these laser systems rely on novel laser architectures \cite{Nubbemeyer:17,Budriunas:17,Balciunas:24} while others utilize advances in spectral broadening and post-compression technology \cite{Mironov_2013,Wheeler_2022}. These near-single cycle pulses offer advantages such as particle acceleration with small, relatively low-power systems and unprecedented repetition rates \cite{Guenot_NaturePhotonics_2017} and the possibility of accelerator optimization with CEP control \cite{Monzac_PRR_2025,angella2025energybunchingsubcycleionization}. 

Since these pulses have spectral bandwidths of 100 nm or more, they present a significant challenge for existing spatio-spectral and spatio-temporal measurement methods. While many metrology methods can, theoretically, cope with such bandwidths, in practice they rely on conventional CMOS sensors and other optical elements which have greatly varying sensitivity over the bandwidth and a small dynamic range. This, coupled with the fact that the pulses themselves often have strongly non-uniform spectra, means that in many cases parts of the pulse are not being reconstructed properly. Upgrading to Indium gallium arsenide (InGaAs) sensors can increase the sensitivity (such as in \cite{Lehotai:25}), but greatly increases the cost of the product.  While some existing techniques  have measured near-single-cycle pulses \cite{Alonso:12} they have often been intrusive, difficult to implement setups that are very sensitive to experimental conditions.

In this paper, we present spatio-spectral and spatio-temporal measurements at the LUCID laser system at the LLC, which provides few-cycle pulses, CEP stable pulses. We show that INSIGHT--which is the closest there is to an industry standard STC measurement device--fails to properly reconstruct the full bandwidth of the laser pulse, due to significant variance in the signal strength over the bandwidth. However, when the pulse is spectrally filtered and then stitched together, the joint measurement succeeds in enhancing the dynamic range of the INSIGHT measurement. To demonstrate the robustness and applicability of this enhancement, we show that the same spectral filtering succeeds in recovering lost spectral information in the IMPALA measurement technique as well as in SRFTS. This simple, inexpensive addition to existing STC metrology methods makes accurate reconstruction of few-cycle pulses a possibility, a critical step to allowing these state-of-the-art facilities to achieve their full potential.

\section{Materials and Methods}

\subsection{LLC Laser System and Experimental Setup}

\begin{figure} [b!] 
   \begin{center}
    \includegraphics[width=\linewidth]{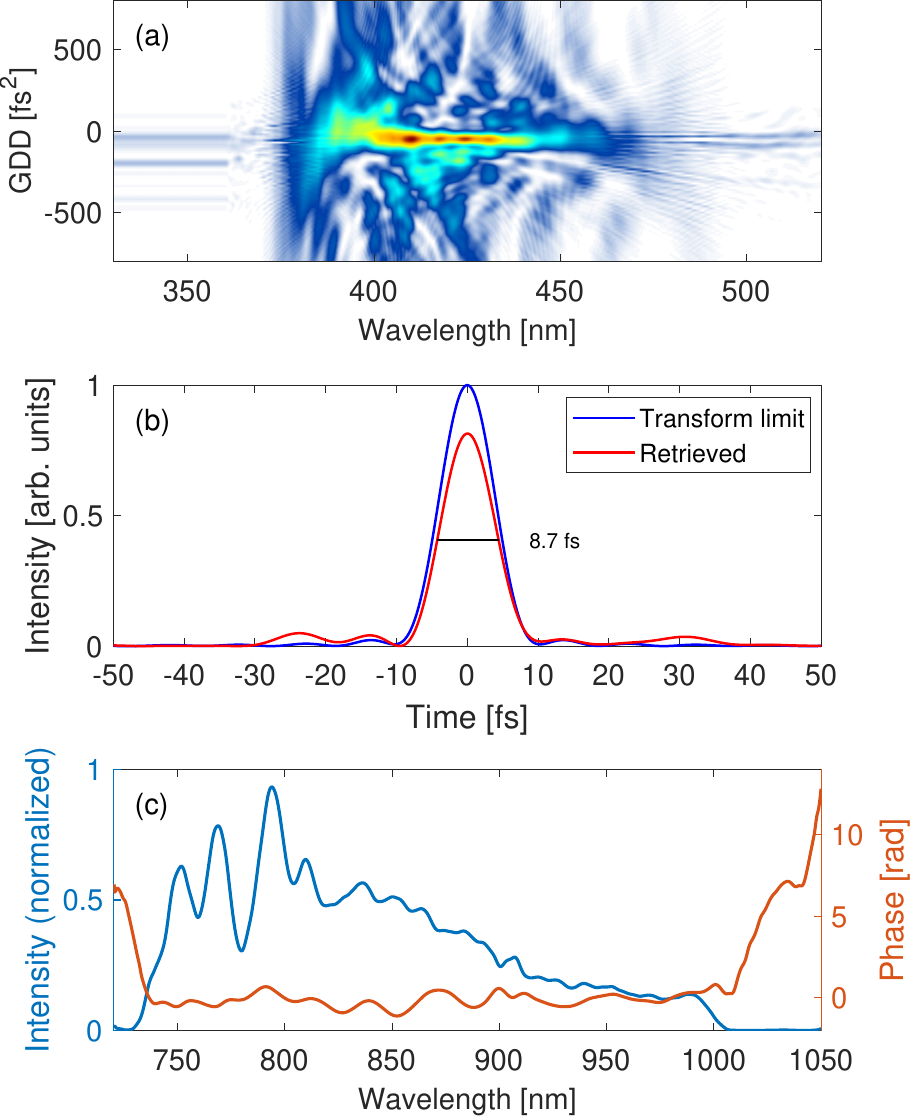}
    \end{center}
    \caption[]{Pulse characterization of the LUCID laser. (a) Dispersion scan measurement. (b) Retrieval of the temporal pulse shape. (c) Retrieved spectrum and spectral phase of the laser. \label{fig:dscan}}
\end{figure}  

\begin{figure*} [t!]
   \begin{center}
    \includegraphics[width=\linewidth]{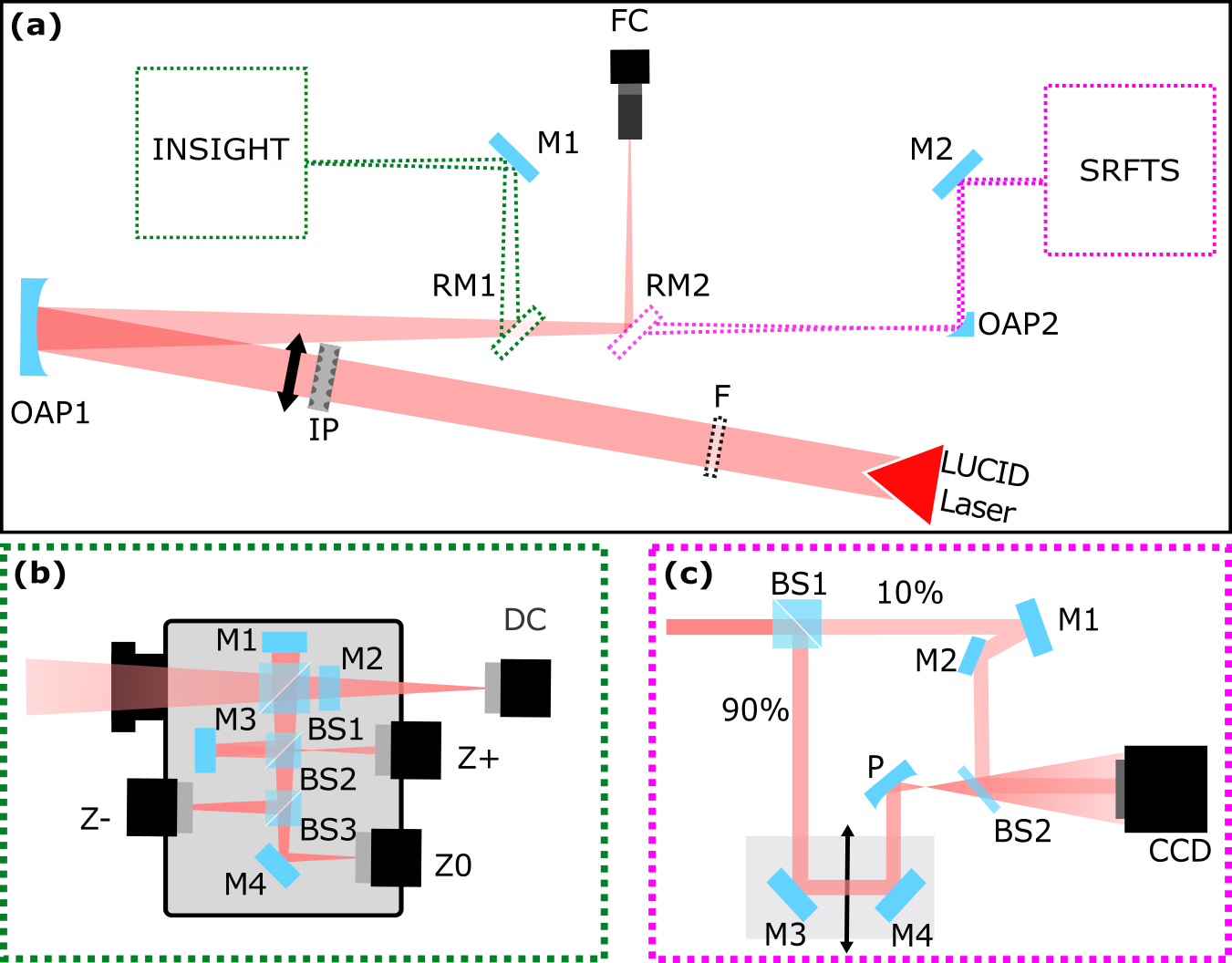}
    \end{center}
    \caption[]{(a) Schematic of the experimental setup. Laser path is shown, as well as the spectral filter (F), IMPALA mask (IP), and focusing off-axis parabola (OAP1). A removable mirror (RM1) channels the laser into the INSIGHT measurement. A different removable mirror (RM2) channels the beam to the focal spot/IMPALA camera (FC). Finally, without both mirrors the beam goes to another off-axis parabola (OAP2) and is steered into the SRFTS measurement. (b) Detailed schematic of the INSIGHT measurement: Mirrors M1 and M2 together with BS1 form a Michelson Interferometer. Mirror M3 and BS2 are used to look after the focal spot on the Z+ camera. BS3 and M4 redirect the beam towards Z- and Z0 cameras. The Direct Camera DC is used for pointing stabilization. (c) Detailed schematic of the SRFTS measurement.BS1 is a 10/90 beamsplitter cube, M denotes flat mirrors, the arrow denotes movement of the piezo stage, P is a 50 mm focusing parabola, BS2 is a 2\% reflection beam splitter and CCD denotes the camera.  \label{fig:Joint_setup}.}
\end{figure*} 

Experiments were carried out using LUCID, the multi-terawatt OPCPA laser system at the Lund Laser Centre \cite{Balciunas2024}, manufactured by Light Conversion. The system delivers 9~fs pulses (intensity-envelope FWHM) with an energy of 250~mJ ($\pm 1\%$) prior to transport, corresponding to 190~mJ on target, at a central wavelength of 850~nm and a 10~Hz repetition rate. 

A dispersive scan (d-scan)  measurement of the pulse can be seen in Fig. \ref{fig:dscan} (a). The dispersive scan works by inserting known thicknesses of glass into the pulse and capturing the second-harmonic generation spectrum. This results in a 2D scan which gives information about the spectrum and the spectral phase and from which the temporal shape of the pulse can be reconstructed. Figure \ref{fig:dscan} (b) shows the retrieved temporal shape of the pulse while figure \ref{fig:dscan} (c) shows the retrieved spectrum and spectral phase. The pulses are compressed in vacuum by means of 16 chirped mirrors with a group delay dispersion (GDD) of $\SI{100}{\femto\second\squared}$ per pair. Aberrations in the spatial profile of the laser can be corrected with an enhanced silver-coated deformable mirror located within the compressor chamber. The short pulse duration of just over 3 laser cycles corresponds to a very wide spectral bandwidth of almost 300~nm, making this laser system highly susceptible to STCs, whose characterization is therefore of great importance. The laser is carrier-envelope phase (CEP) stable due to a DFG based front-end, while the phase drift is detected by an f-2f interferometer located upstream of the compressor and feedbacked to DAZZLER, yielding an RMS CEP stability of 340~mrad. 

The laser's energy can be attenuated either by lowering the pump energy in the last two OPA stages or by inserting a combination of 4 inch OD filters in the beam prior to pulse compression. After the compression stage, all beam transport is vacuum and utilizes enhanced silver-coated mirrors. All beam transport after compression is in the horizontal plane. The beam has a diameter of 50~mm during the vacuum transport. 

For this experiment, the beam was directed to an optical table in air, as shown in Fig. \ref{fig:Joint_setup}. The figure shows an overview of the experimental setup, with the beam path shown starting from the red triangle. Prior to focusing, there is space for the spectral filter (F).  After that, close to the focusing optic, is the insertion point of the IMPALA mask (IP). The beam is then focused by a 4 inch off-axis parabolic mirror (OAP1) with an off-axis angle of 12.8$^\circ$ and an apparent focal length of 1029~mm. From here, the focused pulse can be redirected to the INSIGHT device (green inset) via a removable 2 inch folding mirror (RM3). Alternatively, the focal plane can be imaged using a focus camera with a 10x microscope objective (FC1), by redirecting the converging beam with another 2 inch folding mirror (RM4). This camera also served as the far-field camera for the IMPALA measurement. Finally, the beam can be directed to a second off-axis parabolic mirror (OAP2) with a diameter of 1/2 inch, an off-axis angle of 90$^\circ$ and an apparent focal length of 50.8~mm that is used to re-collimate the beam for diagnosing it using the SRFTS device (magenta inset). The working principles of the aforementioned methods for STC characterization, namely IMPALA, INSIGHT and SRFTS, are detailed in the sections below. All measurements were performed with the unfiltered spectrum and with the insertion of a 2 inch longpass filter with a cut-off wavelength of 850~nm (Thorlabs FGL850S) to increase the spectral dynamic range of the STC reconstruction.

\subsection{INSIGHT Measurement Setup} \label{sec:INSIGHT-method}

The INSIGHT measurement technique \cite{Borot_OE_2018} relies on spatially resolved spectral retrieval of laser pulses via Fourier Transform Spectroscopy, using only beam profiles recorded in the focal plane (Z0) and in two additional planes located a few Rayleigh ranges away from focus (Z+ and Z-). The approach consists of finely scanning the temporal delay between each pulse and an identical reference replica. Once the spectrally resolved beam profiles are obtained both at focus and out of focus, a Gerchberg-Saxton-type algorithm \cite{GS_1972} can be applied to reconstruct the spectrally resolved wavefront.

The method was implemented using the commercial INSIGHT Quadro system (SourceLab). The instrument incorporates, besides the three cameras for Z0, Z- and Z+ beam profiles,  a dedicated pointing‑tracking camera (PTC in Fig. \ref{fig:Joint_setup}) that monitors the pulse direction during the scanning process, enabling compensation for pointing fluctuations of the laser. 
To ensure a broad spectral acceptance, the optical components of the Michelson interferometer were upgraded with broadband optical components (i.e., silver mirrors, including the leakage mirror for the PTC, and broadband filters).

Focusing of the pulses inside the INSIGHT device was achieved directly using the off‑axis parabolic mirror, assisted by two flat 2 inch mirrors (RM3 and M4 in Fig. \ref{fig:Joint_setup}). Neutral‑density filters were positioned at the device entrance to control the incident pulse energy. %Several filters were employed during the measurements, including fixed filters at 780 nm and 850 nm, as well as a variable‑attenuation filter.

The reconstruction of the 3D temporal shape of the pulses starts from the spectral slices of the far-field measured with the INSIGHT for the two measurements: one without any spectral filter and one with an 850 nm filter. 
The spatially-integrated spectrum of a given INSIGHT measurement is reconstructed, and the constant residual noise from the data is estimated. 
Then the constant noise, up to 25 \% in our measurements, is removed from each far-field slice.
Each slice is then normalized to its integrated signal, so that the total intensity of each slice is equal to one.
The normalized datasets are then stitched together, providing increased spectral coverage. 
After stitching, each spectral slice is rescaled according to the corresponding spectral intensity measured independently with a spectrometer. 
The result is an accurate three-dimensional reconstruction of the pulse’s spatio-spectral profile, with improved dynamic range, and respecting the spectrum as measured with the spectrometer. 
Finally,  a Fourier transform is applied along the spectral axis at every point in the XY plane, resulting in the spatio-temporal reconstruction of the pulse.
This step can be performed assuming a flat spectral phase, or with the measured spectral phase from the d-scan device.

\subsection{IMPALA Measurement Setup}

\begin{figure} [h!]
   \begin{center}
    \includegraphics[width=\linewidth]{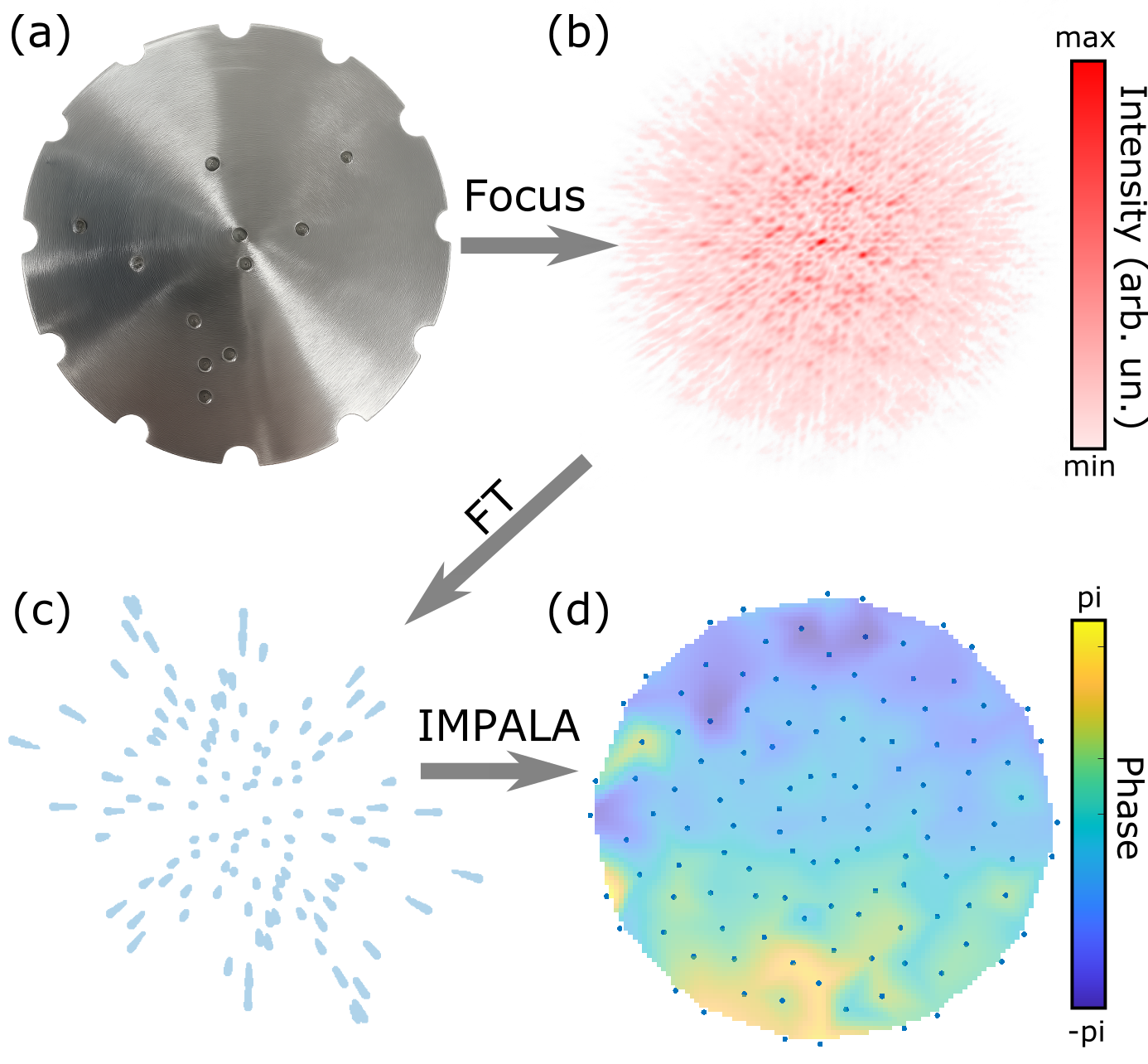}
    \end{center}
    \caption[]{Overview of IMPALA. (a) Pinhole mask designed by genetic algorithm for LLC laser system measurement. (b) Interference speckle pattern that results from passing LLC beam through pinhole mask and then focusing. Red colorscale shows relative intensity. (c) Streak pattern obtained by performing a spatial Fourier transform on the interference speckles. These streaks contain spatially separated chromatic information about the beam. (d) Retrieved wavefront for one particular spectral band of the LLC laser. Colorscale shows phase from $-\pi$ to $\pi$. Blue dots indicate the spatial points at which the wavefront is sampled.\label{fig:Impala_setup} }
\end{figure} 

The principle of IMPALA is to algorithmically, rather than optically, separate out spectral information in the Fourier plane, thus allowing for a minimal experimental setup. The beam is passed through a specially designed pinhole mask, shown in figure \ref{fig:Impala_setup} (a), and focused by the focusing optic used in the experimental setup. The resulting interference speckle pattern, shown in figure \ref{fig:Impala_setup} (b), is captured by an imaging system - in this case the existing focal spot imaging of the experiment. A spatial Fourier transform is then performed on the speckle pattern. In the Fourier plane, this results in a central DC peak and streaks symmetrically distributed around the DC peak, as is shown in figure \ref{fig:Impala_setup} (c). The key to IMPALA is that inside of these streaks, information about the behavior of different wavelengths is spatially separated, with the ``redder" components of the beam closer to the DC peak when compared to the ``bluer" components. The IMPALA algorithm then isolates different spectral bands and performs the GS iterative phase retrieval algorithm to reconstruct a quasi-monochromatic wavefront, shown in figure \ref{fig:Impala_setup} (d), for each spectral band inside of the beam. These wavefronts can then be stitched together, giving the spectrally resolved wavefront of the laser. Additional details of the IMPALA algorithm can be found in Ref. [\cite{Smartsev_OL_2024}].

A specialized mask was designed using the genetic algorithm described in Ref. [\cite{Smartsev_OL_2024}]. This genetic algorithm yields, for a given laser and imaging system and for a given number of holes in the mask, the optimal hole arrangement that yields the minimal overlap of streaks in the Fourier plane. The mask can also be designed to evenly sample the beam with a given number of rotations. In the case of the LLC experiment, a mask with 10 holes in addition to the central hole was used and was designed to sample the beam over 12 rotations. The blue dots in figure \ref{fig:Impala_setup} (d) show the locations along the beam that the wavefront was sampled at. 

Inside of the experimental setup, the IMPALA mask (shown as IP in figure \ref{fig:Joint_setup} (a)) was setup such that the beam passed through the mask immediately prior to hitting the off-axis parabola (OPA1 in figure \ref{fig:Joint_setup} (a)). After focusing, the beam was guided by mirrors (RM2 in figure \ref{fig:Joint_setup} (a)) onto the imaging system (FC in figure \ref{fig:Joint_setup} (a)), which captured the resulting interference pattern. The mask was designed to be removable, such that it would not interfere with the other measurement techniques.

The data was taken with a FLIR BFLY-PGE-31S4M camera that has 2048x1536 pixels, a pixel size of of 3.45 $\mu m$, and a Sony IMX265 sensor. Data was taken with 12 rotations of the IMPALA mask to ensure good spatial coverage of the beam. The data was taken with no filter and with the 850 nm filter. The beam was attenuated to the minimal point at which the interference signal does not saturate the camera during the measurement. There was an attenuation difference of ND 2 between the no filter case and the 850 nm filter case, with the 850 nm filter having the lower attenuation.

%\begin{figure*} [b!]
%   \begin{center}
%   \includegraphics[width=\linewidth]{Figures/Spectral_Stitching_3D.pdf}
%    \end{center}
%    \caption[]{Improvement of spatio-temporal reconstruction of by spectral stitching of INSIGHT measurements. (a) Spatially integrated INSIGHT measurements. Solid line graphs show four different INSIGHT measurements taken without filtering, and with 780 nm, 850 nm, and 900 nm filters. The shaded areas indicate the useful data used in the spectral stitching method. The dashed curve represents an independent spectrum measurement. (b) Highlight of the Fourier-transform-limited (FTL) pulse duration obtained from the spatially-integrated unfiltered INSIGHT measurement and from the independently measured spectrum. (c) Spatio-temporal reconstruction of the unfiltered INSIGHT measurement. (d) Spatio-temporal reconstruction of the stitched data.
%    }
%    \label{fig:spectral_stitching}
%\end{figure*} 
\begin{figure*} [b!]
   \begin{center}
    \includegraphics[width=\linewidth]{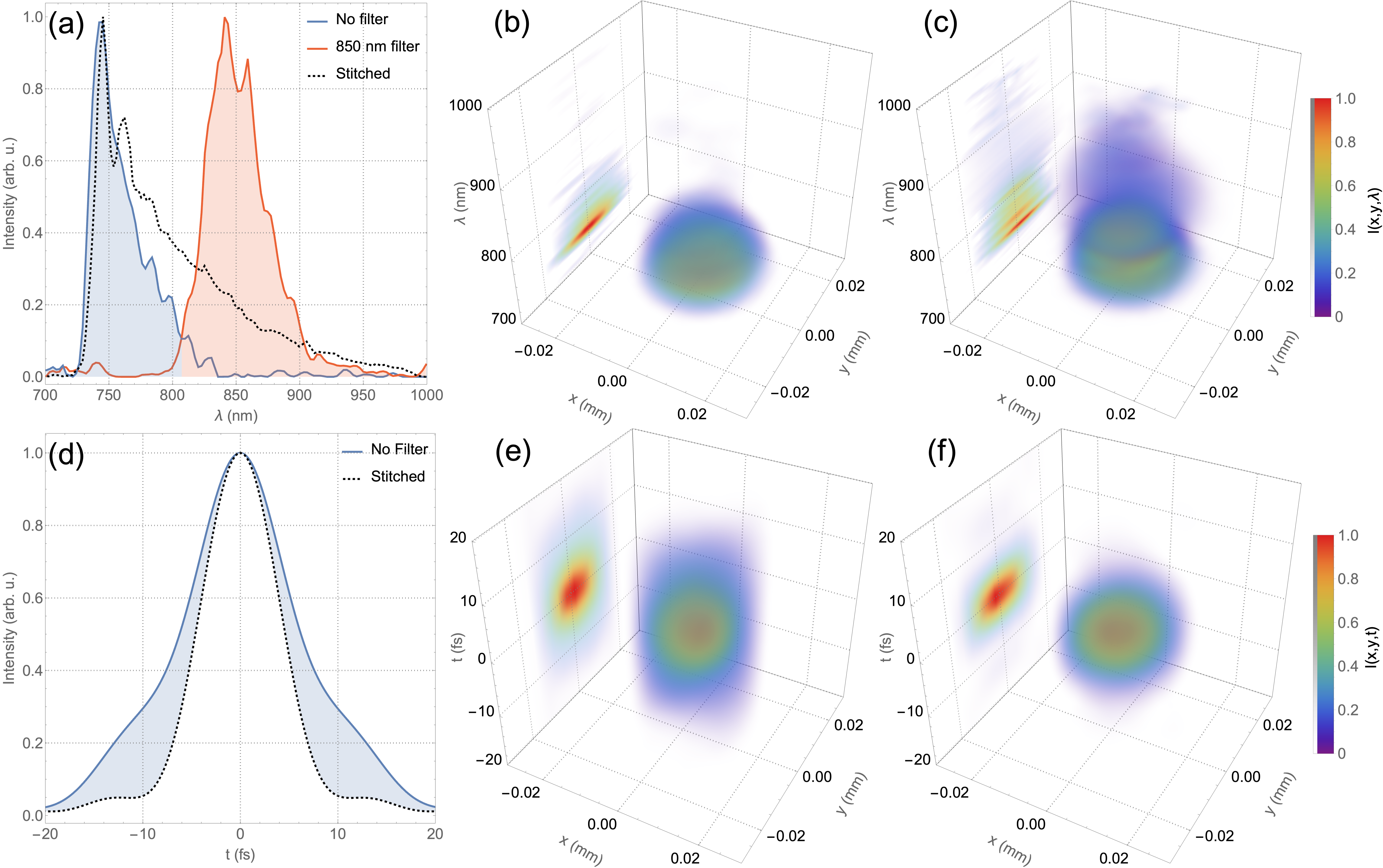}
    \end{center}
    \caption[]{Spectral (a, b, c) and temporal (d, e, f) properties inferred from the measurements:
(a) Spatially integrated INSIGHT spectra from the measurements without filter, and the one with 850 nm cutoff filter (solid lines with shading), and the spectrum measured with a spectrometer (dotted line);
Spatio-spectral reconstruction of the direct INSIGHT measurement (b), and stitched data from the two measurements (c).
(d) Fourier-transform-limited (FTL) pulse duration obtained from the spatially-integrated direct INSIGHT measurement (solid line with shading) and from the spectrometer data (dotted line). 
Spatio-temporal reconstruction using the direct INSIGHT measurement (e), and using the stitched data from the measurements (f).
}
    \label{fig:INSIGHT-stitching}
\end{figure*} 

\subsection{SRFTS Measurement Setup}

Both the INSIGHT and IMPALA measurements were carried out in the focus. In the setup a parabolic mirror was used to focus the beam. To perform the same measurement using the SRFTS device the beam needed to be recollimated. For this purpose, a parabolic mirror (OAP2 in figure \ref{fig:Joint_setup}) was placed behind the focus to obtain a beam of appropriate size. Additional attenuation was performed by routing the beam using a wedge at the entrance of the device.

The SRFTS setup is based on that presented in Ref. \cite{Miranda_OL_2014} and schematically displayed in figure \ref{fig:Joint_setup} (c). The collimated beam is split into two, the unknown and the reference beams, by an uncoated 10/90 beam splitter cube (BS1). The unknown beam propagates undisturbed through the interferometer. The reference beam is first sent to a delay stage, piezzosystemjena PX499SG, where the length of the reference arm is controlled. Then the reference beam is focused by an off-axis parabola (P) with a focal length of 50 mm. The focus can be approximated as a point source, and due to the short focal length, the wavefront becomes sufficiently homogenous upon propagation. The unknown beam and the generated reference beam are recombined, using a 2\% reflection beam splitter (BS2). A CCD camera, FLIR GRASSHOPPER 3 89S6M-C, is used to record the spatial interference for each delay step. Simultaneously, the data set contains the time interferogram for each pixel, allowing reconstruction of the spectrum. Assuming a homogenous reference beam with a spherical phase origin at the focus of the parabola, the phase of the unknown beam and its individual colors can be obtained.

As with the INSIGHT and IMPALA measurements, the SRFTS measurements were carried out without filter and with a chromatic long pass filter, cutting the spectrum below 850nm. The attenuation on the camera was adjusted to optimize contrast while avoiding saturation.

\section{Results and discussion}

\subsection{Enhanced Spectral Bandwidth of INSIGHT}

The large variation in spectral intensity across the spectral range corresponds to a significant variation in the signal-to-noise ratio of the INSIGHT measurements at different spectral ends, impacting the accurate reconstruction of the temporal shape for the pulse.
To address this issue, we introduced the 850 nm spectral cut-off filter to suppress the intense, blue part of the pulse spectrum and to access the spectral components with lower intensity. 
In the presence of the 850 nm filter, the energy of the spectrally clipped pulses entering INSIGHT was increased by removing a neutral density filter from the beam path. 

In figure \ref{fig:INSIGHT-stitching} (a), the two spatially integrated spectra recovered by the INSIGHT device are depicted for the cases of no filter and 850 nm cutoff filter, after constant noise removal (see section \ref{sec:INSIGHT-method} for the step-by-step data processing description).
The spectrum obtained from an independent spectrometer measurement is also shown as a dashed line. 
For a comprehensive spectral description of the two measurements, Fig. \ref{fig:INSIGHT-stitching} (b) depicts the spectrally resolved far field of the INSIGHT measurement without spectral filter. It corresponds to the blue curve in Fig. \ref{fig:INSIGHT-stitching} (a). 
Fig. \ref{fig:INSIGHT-stitching} (c) depicts the spectrally resolved far field corresponding to the stitched reconstruction using the spectral intensity represented with dashed line in Fig. \ref{fig:INSIGHT-stitching} (a).

To understand the effects on the temporal reconstruction of the pulses, we assumed a spectrally flat phase and computed the pulse duration for the case of the INSIGHT measurement without spectral filter and for the case of the stitched data, as depicted in figure \ref{fig:INSIGHT-stitching} (d). 
A significant reduction in the Fourier-limited pulse duration is observed in the case of the spectrum obtained from the direct INSIGHT measurement (blue line) compared to the spectrum recorded with the spectrometer (dotted line): from 12.4 fs to 9.3 fs FWHM, corresponding to a reduction by a factor of 1.33.

Fig. \ref{fig:INSIGHT-stitching} (e) and Fig. \ref{fig:INSIGHT-stitching} (f), depict the spatially resolved temporal intensity distribution in the focus reconstructed from the direct INSIGHT measurement (Fig. \ref{fig:INSIGHT-stitching} (c)) and from the two  measurements stitched. 
A significant reduction in the focal volume is observed in the case of the stitched measurement, as expected for a broader bandwidth and a more realistic reconstruction. 

The results show that the INSIGHT device, even when adapted with silver mirrors and broadband optics, failed to accurately reconstruct the LUCID laser, due to the limitation of the sensor. 

The significant discrepancy between the spectrometer-given pulse duration and the pulse duration obtained from the no-filter measurement also impacts the value of the peak intensity. 
Integrating the curves in Fig. \ref{fig:INSIGHT-stitching} (d), and assuming the same energy for the pulse in the two cases, the peak intensity retrieved from the spectrometer measurement is 55 \% larger than the one from the direct INSIGHT measurement. This value is similar (57\%) also when considering the full 3D integration of the reconstructed pulses in Fig. \ref{fig:INSIGHT-stitching} (e) and (f).

\subsection{Enhancing Spectral Bandwidth of IMPALA}

\begin{figure*} [t!]
   \begin{center}
    \includegraphics[width=\linewidth]{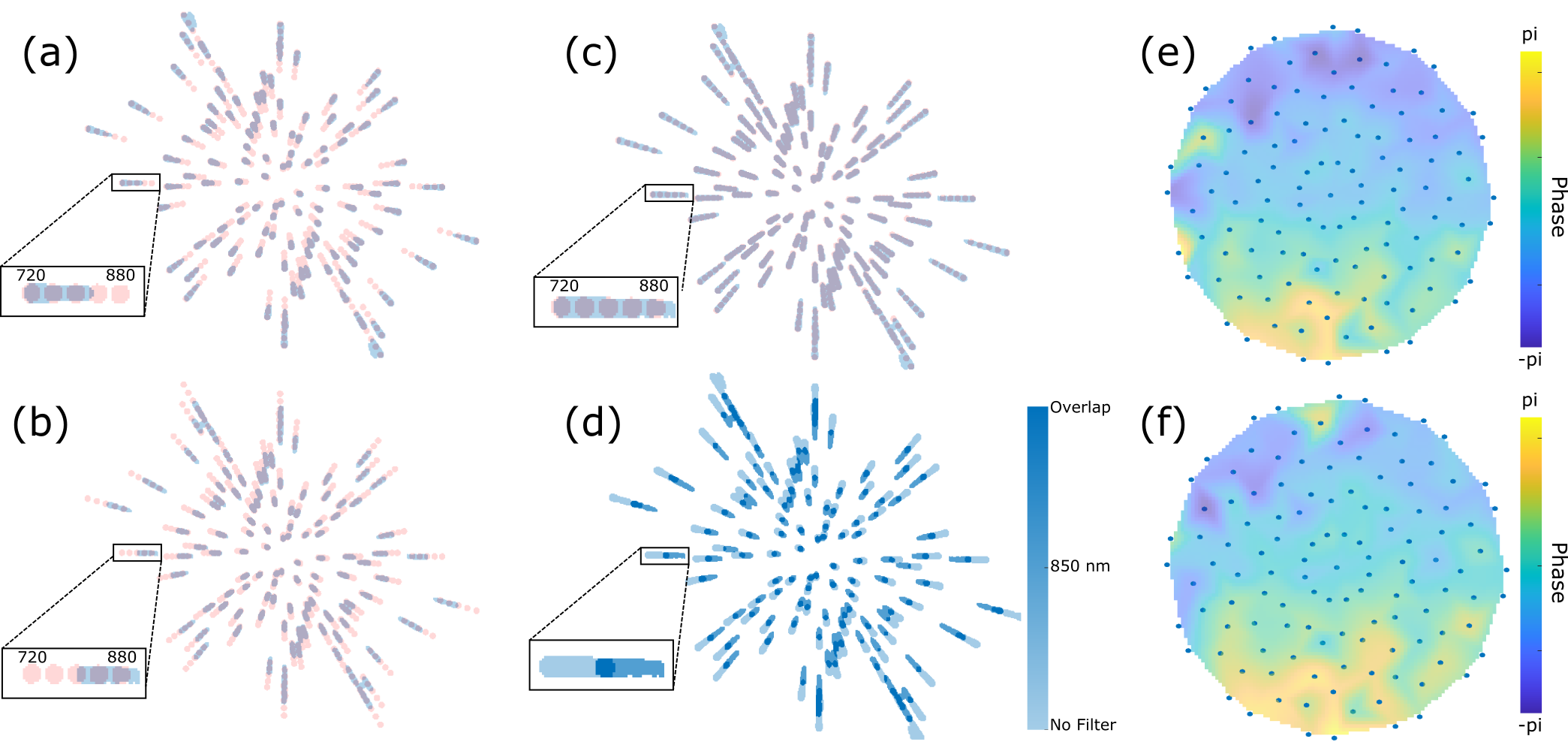}
    \end{center}
    \caption[]{Impact of dynamic range enhancement on IMPALA data. (a) IMPALA streaks (blue) for LLC laser without any spectral filtering as well as the simulated quasi-monochromatic streaks centered at 720 nm, 760 nm, 800 nm, 840 nm, and 880 nm (shown in pink). The insert shows an enlarged example of a streak. (b) IMPALA streaks for LLC laser with 850 nm spectral filter in place. (c) IMPALA streaks obtained by summing the results of the no filter and 850 nm filter cases. All areas with signal are normalized to the same value. (d) IMPALA streaks obtained by summing the results of the no filter and 850 nm filter cases. The blue colorbar shows the places where just one signal is present or where overlapping signals are present. (e) Retrieved 760 nm wavefront for the LLC laser. Colorscale shows phase from $-\pi$ to $\pi$. Blue dots indicate the spatial points at which the wavefront is sampled. (f) Retrieved 840 nm wavefront.    \label{fig:Impala_data} }
\end{figure*} 

For the IMPALA measurement, the interference speckle patterns were Fourier transformed to obtain the speckle patterns, as described above. Figure \ref{fig:Impala_data} (a) shows the experimentally obtained speckles for the LLC beam with no spectral filters present and for one particular orientation of the IMPALA mask. The speckles shown have been normalized to have a value of unity where signal is present and a value of zero where no signal is present. In addition, the figure shows the simulated quasi-monochromatic speckles centered at different wavelengths, ranging from 720 nm (the most distant from the center) to 880 nm (closest to center) in 40 nm steps. The insert shows an enlarged example of one streak, for emphasis. As can be seen in the insert, the recovered spectrum of the streaks spans about 720nm-820nm. From the measured spectrum of the LLC laser system in figure \ref{fig:dscan} (c), it is clear that, as with INSIGHT, a significant portion of the spectrum is lost due to insufficient signal being present at the ``redder" wavelengths of the laser. 

When the 850 nm filter is introduced into the laser path, the IMPALA speckles are significantly altered. As can be seen in figure \ref{fig:Impala_data} (b), the spectral range of the speckle is now about 800 nm - 900 nm, representing a significant redshift over the previous signal. This is due to the fact that with the 850 nm filter in place, ND 2 of attenuation was removed, thus allowing the ``redder" part of the beam to have a strong enough signal to be detected. The signal above 900 nm was still too weak to detect. 

Figure \ref{fig:Impala_data} (c) shows result of summing the experimental no filter streaks and 850 filter streaks and then normalizing the result in the same way as in figures \ref{fig:Impala_data} (a,b). As can be clearly seen, with the addition of the two results the signal now spans from 720 nm - 900 nm, covering a far larger span than each individual measurement contained. This is further flushed out in figure \ref{fig:Impala_data} (d), which again shows the summed data of the no filter streaks and the 850 filter streaks. Here, however, rather than normalizing to unity, the colorbar in the figure shows the areas in which only the no filter data is present, only the 850 nm filter data is present, and where both datasets overlap. 

It is important to note that the no filter data and the 850 nm filter data were processed in exactly the same way and no reorienting was introduced in order to align the two different streak patterns. Simply by numerically adding the results together, the streaks naturally align in a near-perfect manner. This clean alignment indicates that introducing the spectral filter effectively expands the range of the measurement while introducing minimal distortion to the data itself.  

Figure \ref{fig:Impala_data} (e) shows the retrieved monochromatic wavefront at 760 nm, obtained using IMPALA. The colorscale shows phase from $-\pi$ to $\pi$ and the blue dots indicate the spatial points at which the wavefront is sampled. Figure \ref{fig:Impala_data} (f) shows the retrieved monochromatic wavefront but at 840 nm. The ability to retrieve wavefronts from a broad range of frequencies, beyond the range of that afforded by the measurement with no filters, shows the advantage of the spectral enhancement method.

\subsection{Enhancing Spectral Bandwidth of SRFTS}
\begin{figure}[h]
    \centering
    \includegraphics[width=1\linewidth]{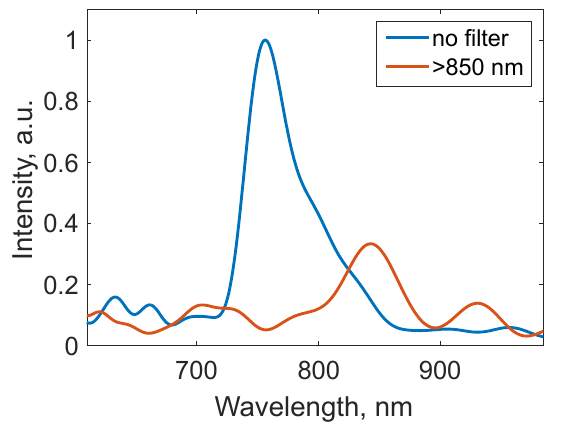}
    \caption{Retrieved spectrum from the SRFTS measurement with no filter (blue) and long-pass filter at 850 nm (red).}
    \label{fig:SpectraSRFTS}
\end{figure}

SRFTS measurements were taken without and with the spectral filter. Figure \ref{fig:SpectraSRFTS} shows the spectrum retrieved from the two measurements. The spectra were obtained by retrieving the spectrum in each pixel through Fourier transform spectrometry, and then spatially integrated to obtain the spectrum of the full beam. Instead of normalizing each spectrum to unity, only the unfiltered spectrum has been normalized, while the other has been normalized and then matched to that noise level. When comparing the two measurements, it is clear that the case without filter (blue) yields a peak at 754 nm, which is shifted to 846 with the filter (red). Introducing a filter that mostly attenuates the blue part of the spectrum allows retrieval of red wavelengths that were not retrievable without the filters. The retrieved spectrum of the laser spans approximately from 720 nm to 850 nm, with a clear triangular shape. Introducing the chromatic filter shifts this to 800-880 nm. 

\section{Conclusion}

The centrality of the CMOS sensor to many spatio-temporal and spatio-spectral measurement methods is a significant barrier in the accurate reconstruction of few-cycle pulses in the near-infrared regime. As was shown, all three methods tested, INSIGHT, IMPALA, and SRFTS, failed to accurately reconstruct the full spectrum of the pulse and therefore missed the spatio-spectral behavior of those wavelengths. The solution shown on all three methods, the introduction of chromatic filters, succeeded in overcoming this limitation by allowing the wavelengths in which the CMOS is less sensitive to be amplified more and, therefore, to cross the signal to noise threshold of the detector. This solution can be further expanded by introducing more spectral filters, allowing the stitching of even larger bandwidths. 

Another possible solution would be to manufacture spectral filters that pre-compensate for the specific CMOS response so that, by placing them before the sensors, the CMOS's spectral response would be flattened. This, however, would be a solution tailored to a specific system while the solution used in this work is easily implementable in any laser system. By enabling the extension of existing metrology methods to few-cycle pulses, this work lays the ground for the accurate reconstruction of those pulses. This ability is both critical to the avoidance of unwanted spatio-temporal/spatio-spectral effects that could limit the quality of the pulse as well as to the accurate implementation of such effects for advanced spatio-temporal shaping of the laser pulse.

\begin{center}\textbf{Author Contributions}\end{center}

The joint experiment was conceptualized and planned by D.U., A.L., O.L., C.Ar., and V.M.. The manuscript was prepared under the supervision of A.L. with input from all the authors. All authors except  V. M. participated in the experiment. All authors participated in preparatory work for the experiment. The LASERLAB grant was obtained by D.U.. Funding for the ELI-NP team was obtained by D.U. who also supervised the team; Funding for the LLC team was obtained by O.L. and C.Ar. who also supervised the team;  Funding for the Weizmann team was obtained by V.M. and the team was supervised by A.L. and V.M..

\begin{center}\textbf{Acknowledgment}\end{center}

\noindent The joint experiment received funding from LASERLAB-EUROPE (grant agreement no. 871124, European Union’s Horizon 2020 research and innovation programme)

\noindent  ELI-NP team was funded by Romanian Ministry for Research, Innovation and Digitalization through ELI-NP IOSIN program,  through project ELI-RO/DFG/2025_013 IATP-NP 2.0 and ELI-RO/AMIPulse funded by Institute of Atomic Physics Romania, and through Nucleu program PN 23 21 01 05. 

\noindent The Weizmann team was funded by the Schwartz/Reisman Center for Intense Laser Physics, the Benoziyo Endowment Fund for the Advancement of Science, the Israel Science Foundation, Minerva, Wolfson Foundation, the Schilling Foundation, R. Lapon, Dita and Yehuda Bronicki, and the Helmholtz Association.

\noindent The LLC team was supported by the Swedish Research Council (VR 2019-04784, 2024-05698), the Knut and Alice Wallenberg Foundation (KAW 2019.0318, 2020.0111) and the European Union's Horizon Europe research and innovation programme (Grant Agreement No. 101073480).

\noindent The authors thank Dr.~Slava Smartsev for constructive conversations about the IMPALA measurement. The authors thank Prof.~Claes-Göran Wahlström and Dr.~Ann-Kathrin Raab for assistance in the planning and execution of the experiment. 

\begin{center}\textbf{References}\end{center}
\vspace{-28pt}
\renewcommand{\refname}{}
\bibliographystyle{hplse}
\bibliography{Refs.bib}

\vspace*{4mm}

\end{document}